\documentclass[11pt]{article}
\usepackage{graphicx}
\usepackage{dcolumn}
\usepackage{array, longtable}
\usepackage{epsfig}
\usepackage{amsmath}
\usepackage{colordvi}
\usepackage{hhline}
\usepackage{tabularx,pifont,multirow}

\newcommand{\BABARPubYear}    {06}

\newcommand{\BABARConfNumber} {043}
\newcommand{\SLACPubNumber} {11995}
\newcommand{\LANLNumber} {0000}

\long\def\inst#1{\par\nobreak\kern 4pt\nobreak
    {\it #1}\par\vskip 10pt plus 3pt minus 3pt}

\input babarsym
\def\pom {\ensuremath{\pm}\xspace}

\def\Imm       {\ensuremath{\Im m}}
\def\Ree       {\ensuremath{\Re e}}

\def\Kstarone   {\ensuremath{K^*(892)}\xspace}

\def\mell       {\ensuremath{\ell}\xspace}

\newcommand{\vomega}{\boldsymbol{\omega}}
\newcommand{\jchan}{{b}}

\begin{document}

%\begin{flushleft}
%Babar Analysis Document \# 1404,   Version 16 \\
%Charmon-04/03
%\end{flushleft}

%\vspace{-1.75cm}

\begin{flushright}
\babar-CONF-\BABARPubYear/\BABARConfNumber \\
%\babar-PUB-\BABARPubYear/\BABARPubNumber \\
SLAC-PUB-\SLACPubNumber \\
hep-ex/\LANLNumber \\
July 2006 \\
\end{flushright}

\vfill

\begin{center}{\large \bf 
Measurement of Decay Amplitudes of \B \to (\ccbar) \Kstar with an 
angular analysis, for (\ccbar)=\jpsi, \psitwos and \chicone.
}

\bigskip

\begin{center}
\large The \babar\ Collaboration\\
\mbox{ }\\
\today
\end{center}

\bigskip 
\bigskip \bigskip
\bigskip \bigskip

\large {\bf Abstract}
\end{center}

We perform the first three-dimensional measurement of the amplitudes of
$\B\to\psitwos\Kstar$ and $\B\to\chicone\Kstar$ decays and update our
previous measurement for $\B\to\jpsi\Kstar$.
We use a data sample collected with the \babar\ detector at the \pep2\
storage ring, representing 232 million produced \BB\ pairs.
The longitudinal polarization of decays to the $1^{++}$ \chicone meson
 together with a \Kstar meson, is found to be larger than that for the
 decay to the $1^{--}$ $\Psi$ mesons.
No direct \CP-violating charge asymmetry is observed.

\bigskip \bigskip

\begin{center}
Submitted to the 33$^{\rm rd}$ International Conference on High-Energy Physics, ICHEP 06,\\
26 July---2 August 2006, Moscow, Russia.
\end{center}

\vspace{1.0cm}
\begin{center}
{\em Stanford Linear Accelerator Center, Stanford University, 
Stanford, CA 94309} \\ \vspace{0.1cm}\hrule\vspace{0.1cm}
Work supported in part by Department of Energy contract DE-AC03-76SF00515.
\end{center}

\newpage
\setcounter{footnote}{0}

\begin{center}
\small

The \babar\ Collaboration,
\bigskip

%% author list as of 01-Jul-2006 (596 authors)
%
{B.~Aubert,}
{R.~Barate,}
{M.~Bona,}
{D.~Boutigny,}
{F.~Couderc,}
{Y.~Karyotakis,}
{J.~P.~Lees,}
{V.~Poireau,}
{V.~Tisserand,}
{A.~Zghiche}
\inst{Laboratoire de Physique des Particules, IN2P3/CNRS et Universit\'e de Savoie,
 F-74941 Annecy-Le-Vieux, France }
% ...

\end{center}\newpage

\section{Introduction}
\label{sec:Introduction}

\B decays to charmonium-containing final states  (\jpsi, \psitwos, \chicone) \Kstar are of interest for the precise
measurement of $\sin 2 \beta$, as for the similar decay $\B \to \jpsi
\Kz$.  
Here,  the final state consists of two vector particles, the \Kstar meson and the  charmonium  meson  : the $L=1$
and $L=0,2$ states have different \CP eigenvalues and the related
dilution of any \CP violation must be taken into account in the
measurement of $\sin 2 \beta$.
The amplitude for  longitudinal polarization of the two
vector mesons is $A_0$. 
There are two amplitudes for polarizations of the vector mesons
transverse to the decay axis: $A_\parallel $ for parallel polarization
of the two vector mesons and $A_\perp$ for their perpendicular polarization.
Only the relative amplitudes are measured here, so that 
$|A_0|^2 + |A_\parallel |^2 + |A_\perp|^2 = 1$.
Previous measurements of the amplitudes by the 
CLEO \cite{Jessop:1997jk},
CDF \cite{Affolder:2000ec},
\babar\ \cite{Aubert:2004cp}
and Belle \cite{Itoh:2005ks}
collaborations for the $\B\to\jpsi\Kstar$ channels are all compatible with
each other, and with a \CP-odd intensity fraction $|A_\perp|^2$  close to
0.2.

Factorization is a framework that allows the description of heavy 
meson decays by assuming that a weak decay matrix element can be
described as the product of two independent hadronic currents.
In the case of heavy quarks present in the final state,
the validity of the factorization hypothesis can be questioned.
Factorization predicts 
that the phases of the decay amplitudes are the same (modulo $\pi$). 
\babar\ has observed \cite{Aubert:2004cp,Aubert:2001pe}
a significant departure from this prediction.
The factorization-suppressed decay $\B\to\chiczero\Kpm$ has also been
observed \cite{Aubert:2003vc,Garmash:2004wa} with a branching fraction
of the same order of magnitude as 
that of the factorization-allowed $\B\to\chicone\Kpm$ ,
while the decay to \chictwo, 
predicted to have non-factorizable contributions comparable to those
for decays to \chiczero \cite{Colangelo:2002mj}, is actually not
observed \cite{Aubert:2005vw}.

Precise measurements of the branching fractions of these decays are
now available \cite{Aubert:2004rz} to test the theoretical description
of the non-factorizable contributions \cite{Chen:2005ht}, but
polarization measurements are also needed.
In particular measurements for \psitwos and \chicone, compared to
that of \jpsi, would discriminate the mass dependence from the quantum
number dependence, due to the different effective Hamiltonian matrix
element that describes charmonium production from vacuum under the
factorization hypothesis, and to the different non-factorizable
contributions \cite{Chen:2005ht}.
CLEO has measured the longitudinal polarization of 
 \B\to\psitwos\Kstar decays to be $|A_0|^2 =0.45 \pm 0.11 \pm 0.04 $ 
\cite{Richichi:2000ca}.
Belle has studied \B\to\chicone\Kstar
decays and obtained $|A_0|^2 =0.87 \pm 0.09 \pm 0.07$ \cite{Soni:2005fw}.

\B decays to charmonium $K^{(*)}$ provide a clean environment for the
measurement of CKM angles because one tree amplitude  dominates
the decay. 
Very small direct \CP-violating charge asymmetries are
expected in these decays : the observation of a sizeable, significant
signal would be a smoking gun for the presence of new physics.
No such signal has been found  \cite{Aubert:2004rz}.
London {\em et al.} have suggested that several amplitudes with both
different electro-weak phases and different strong phases must be
present to create a charge asymmetry in a simple branching fraction
measurement, while an angular analysis of vector-vector decays can
detect charge asymmetries even in the case of vanishing strong phase
difference \cite{London:2000zi}.
Belle has looked for, and not found, such a signal  \cite{Itoh:2005ks}.

In this paper we describe the amplitude measurement of \B \to (\ccbar)
 \Kstar with an angular analysis, for (\ccbar)=\jpsi, \psitwos and
 \chicone, using a selection similar to that of
Ref. \cite{Aubert:2004rz} and described below, and a fitting method similar to that of
Ref. \cite{Aubert:2004cp}.
$\Psi$ candidates (any of the $1^{--}$ charmonia; i.e. \jpsi or \psitwos) 
are  reconstructed in their decays to   $\mell\mell$,
where \mell represents an electron or a muon, and \chicone candidates to $\jpsi\g$.
Decays to the flavor-eigenstates 
$\Kstarz\to\Kpm\pimp$, 
$\Kstarpm\to\KS\pipm$ and
$\Kstarpm\to\Kpm\piz$
are used.
The relative strong phases are known to have a two-fold
ambiguity when measured in an angular analysis alone.
In contrast with earlier publications
\cite{Jessop:1997jk,Affolder:2000ec,Aubert:2001pe} we use here the set
of phases predicted by Suzuki \cite{Suzuki:2001za} using arguments
based on the conservation of the $s$ quark helicity in the decay of
the $b$ quark.
We have confirmed experimentally this prediction by the study of
the variation with $K\pi$ invariant mass of the phase difference
between the \Kstarone amplitude and a  non-resonant  $K\pi$ S-wave amplitude
\cite{Aubert:2004cp}.

\section{\boldmath The \babar\ detector and dataset}
\label{sec:babar}

The data, collected with the \babar\ detector at the \pep2\ asymmetric
\epem storage ring, represent 232 million produced \BB\ pairs, corresponding to
an on-resonance integrated luminosity of about 209 \invfb.

The \babar\ detector is described in detail elsewhere~\cite{detector}.
Charged-particle tracking is provided by a five-layer silicon
vertex tracker (SVT) and a 40-layer drift chamber (DCH). 
For charged-particle identification (PID), ionization energy loss in
the DCH and SVT, and Cherenkov radiation detected in a ring-imaging
device (DIRC) are used.
Photons are identified by the electromagnetic calorimeter
(EMC), which comprises 6580 thallium-doped CsI crystals. 
These systems are mounted inside a 1.5-T solenoidal
superconducting magnet. 
Muons are identified in the instrumented flux return (IFR), composed
of resistive plate chambers and layers of iron that return the
magnetic flux of the solenoid.

We use the GEANT~\cite{geant} software to simulate interactions of particles
traversing the detector, taking into account the varying
accelerator and detector conditions. 

\section{\boldmath Event Selection}
\label{sec:Selection}

Event pre-selection is performed in the same way as in  Ref. \cite{Aubert:2004rz}.
Multihadron events are selected by demanding a minimum of three
reconstructed charged tracks in the polar-angle range $0.41 <\theta <
2.54$\rad, where $\theta$ is defined in the laboratory frame. Charged
tracks must be reconstructed in the
DCH and are required to originate within 1.5\cm of the beam in the
plane transverse to it and within 10\cm of the beamspot along the beam
direction.
Events are required to have a primary vertex within 0.5\cm of the
average position of the interaction point in the plane transverse to
the beamline, and within 6\cm longitudinally.
Charged tracks are required to include at least 12 DCH hits and to
have a transverse momentum $\pt > 100 \mevc$.
Photon candidates are required to have a minimum energy of 30 MeV, to
have a lateral energy profile compatible with that of an
electromagnetic shower, and to be in the fiducial volume of the EMC,
$0.41 < \theta < 2.41$ rad.
Electron candidates are selected using information from the EMC, the ratio of the
energy measured in the EMC to the momentum measured by the tracking
system, the energy loss in the drift chamber, and the Cherenkov angle
measured in the DIRC.
Electrons are also required to be in the fiducial volume $0.41 <
\theta < 2.41$ rad.
Muon candidates are selected using information from the EMC (energy deposition
consistent with a minimum ionizing particle) and the distribution of hits in the
RPC. Muons are required to be in the fiducial volume $0.3 < \theta < 2.7$ rad.
We select charged kaon and pion candidates using information from the
energy loss in the SVT and DCH, and the Cherenkov angle measured in the DIRC. Kaon
candidates are required to be in the fiducial volume $0.45 < \theta < 2.45$ rad.

\B candidates are selected in a similar way as in Ref. \cite{Aubert:2004rz}.
The \jpsi candidates are required to have an invariant 
mass $2.95 < m_{\epem} < 3.14$ \gevcc or $3.06 < m_{\mumu} < 3.14$ \gevcc for decays to \epem 
and to \mumu respectively. 
The \psitwos candidates are required to have invariant masses 
$3.44 < m_{\epem} < 3.74$ \gevcc or $3.64 < m_{\mumu} < 3.74$ \gevcc. 
Electron candidates are combined with photon candidates in order to
recover some of the energy lost through bremsstrahlung.
In the \chicone reconstruction, 
the associated \g has to satisfy shower shape requirements 
% LAT $<$ 0.8, $A_{42}<$0.15 
and has to
have an energy greater than 150 \mev.
The \chicone candidates are required to satisfy 
$350 < m_{\ellell \g} - m_{\ellell} < 450$ \mevcc.
The \piz \ra \gaga candidates are required to satisfy $113 < m_{\gaga} < 153$ \mevcc. 
%Both photons have to satisfy LAT $<$ 0.8. 
The energy 
of the soft photon has to be greater than 50 \mev, and the energy of the hard 
photon has to be greater than 150 \mev. 
The \KS \ra \pip \pim  candidates are required to satisfy $489 < m_{\pipi} < 507$ \mevcc. 
In addition, the \KS flight distance from the $\Psi$ vertex must be larger than 3 standard deviations.
The \Kstarz and \Kstarp candidates are required to satisfy 
$796 < m_{\kaon \pi} < 996$ \mevcc and $792 < m_{\kaon \pi} < 992$ \mevcc, 
respectively. 
In addition, due to the presence of a large background of low-energy non-genuine
 \piz's, the cosine of the angle $\theta_{K^*}$ between the \kaon
 momentum and the \B momentum in the \Kstar rest frame has to be less
 than 0.8  for $\Kstar \to\Kpm\piz$.
For events which reconstruct to $\B_a \to V \Kstar_a$ and $\B_b \to V
\Kstar_b$ modes, with $\Kstar_a$ decaying to \piz and $\Kstar_b$
decaying to \pipm, the $\B_a$ candidate is discarded, as $\pipm
\to\piz$ is observed to be the dominant source of cross-feed.

\begin{figure}
\begin{center}
  \includegraphics[width=0.32\textwidth]{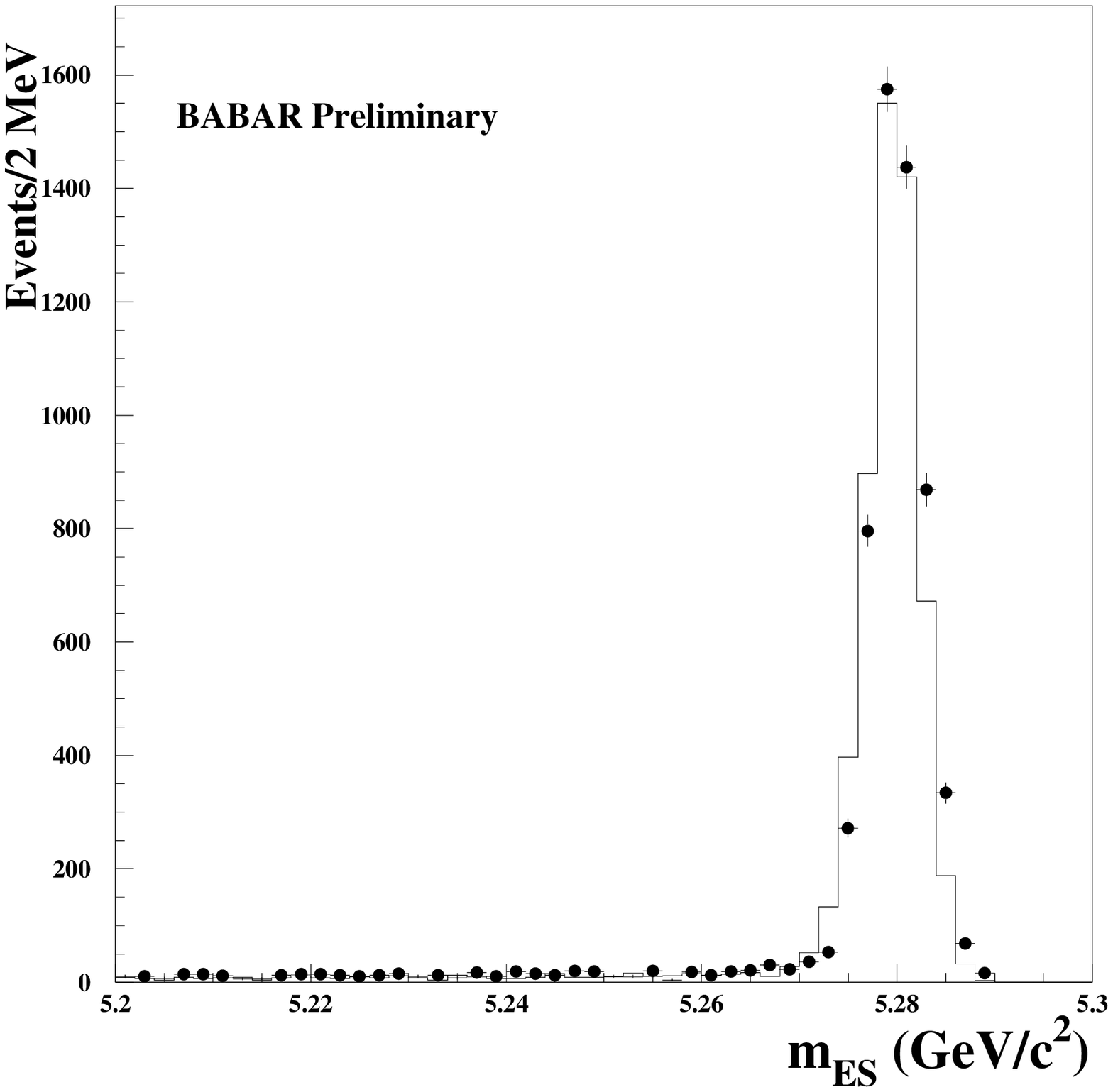}
  \includegraphics[width=0.32\textwidth]{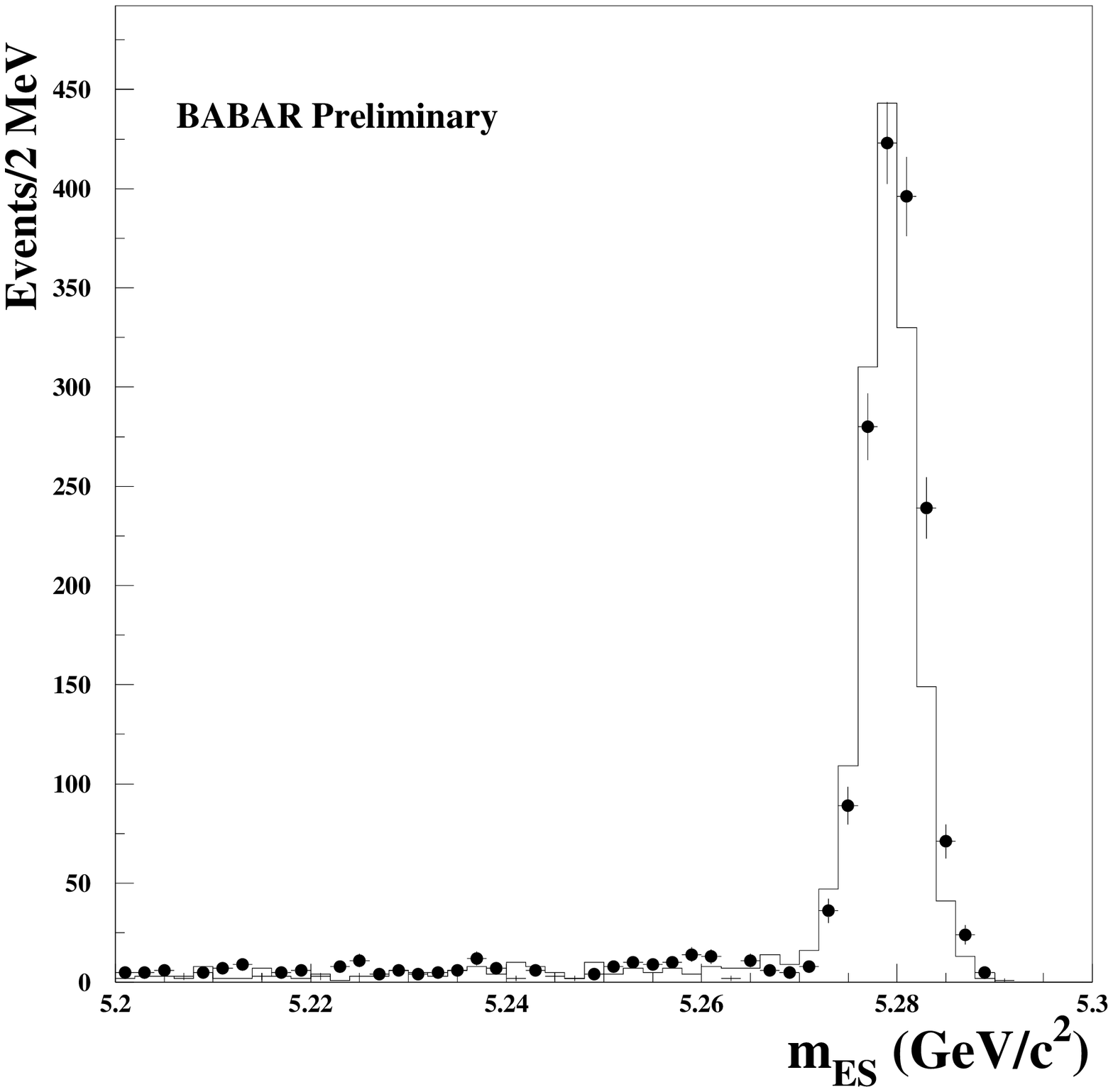}
  \includegraphics[width=0.32\textwidth]{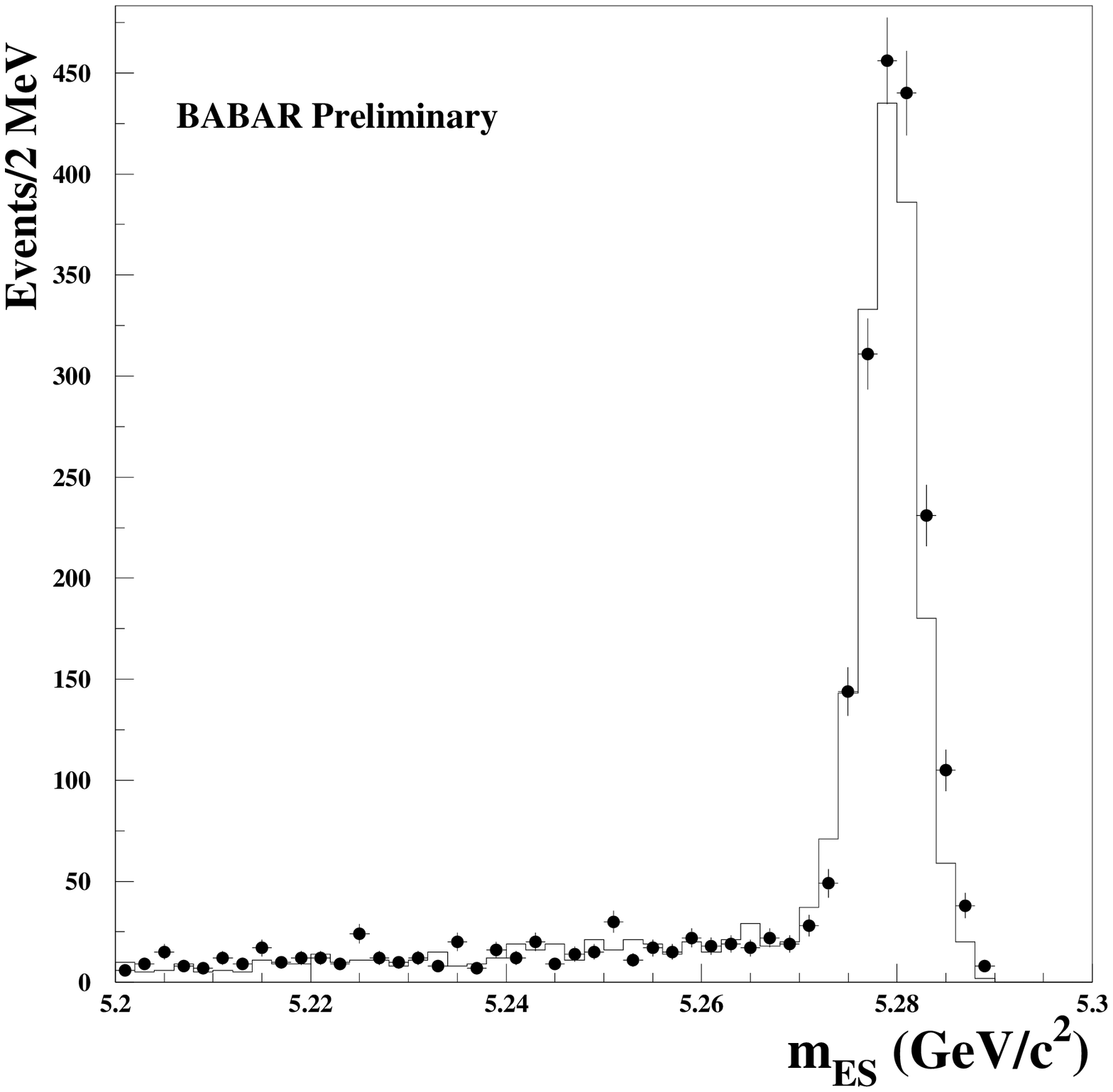}
  \includegraphics[width=0.32\textwidth]{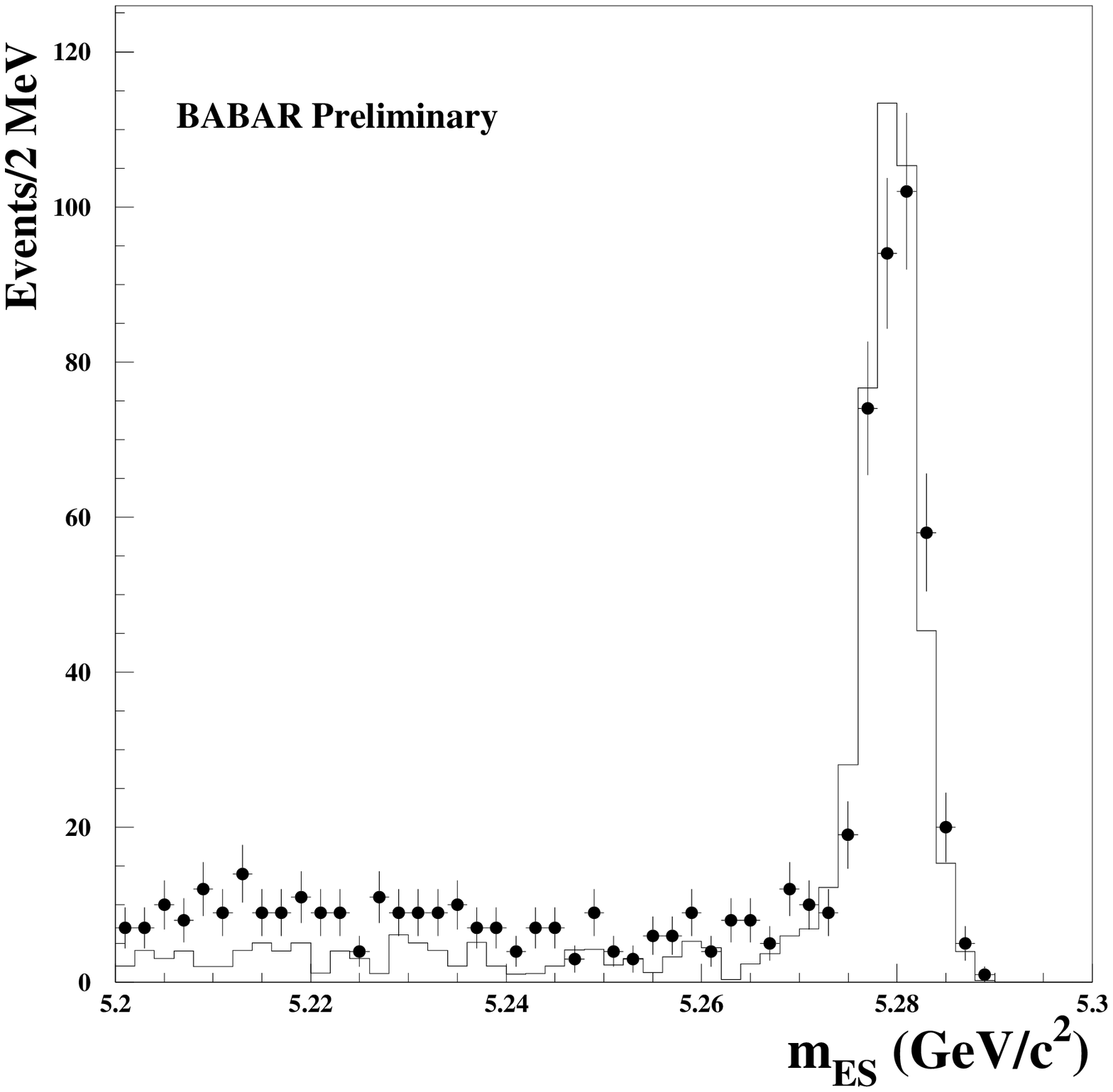}
  \includegraphics[width=0.32\textwidth]{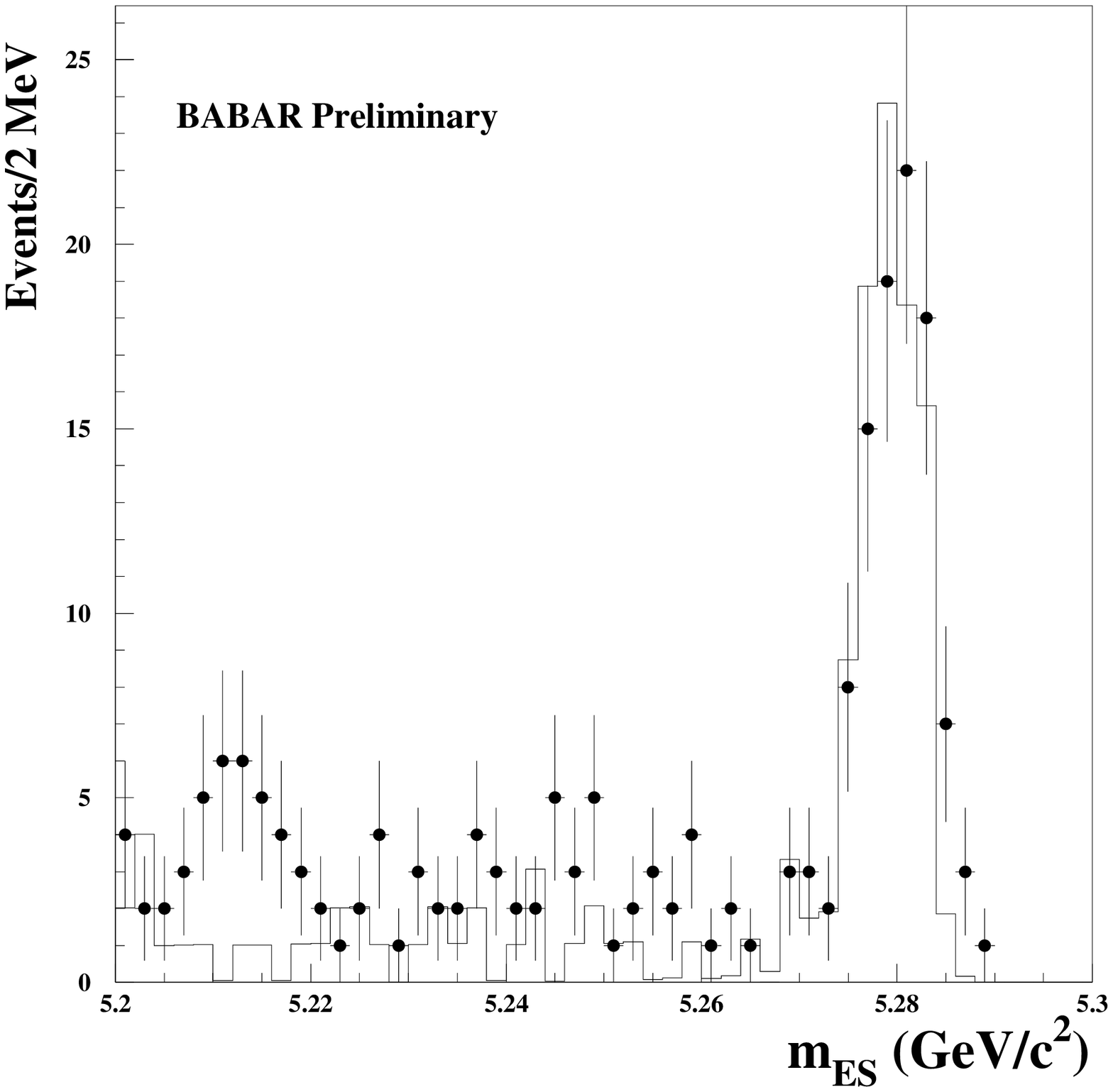}
  \includegraphics[width=0.32\textwidth]{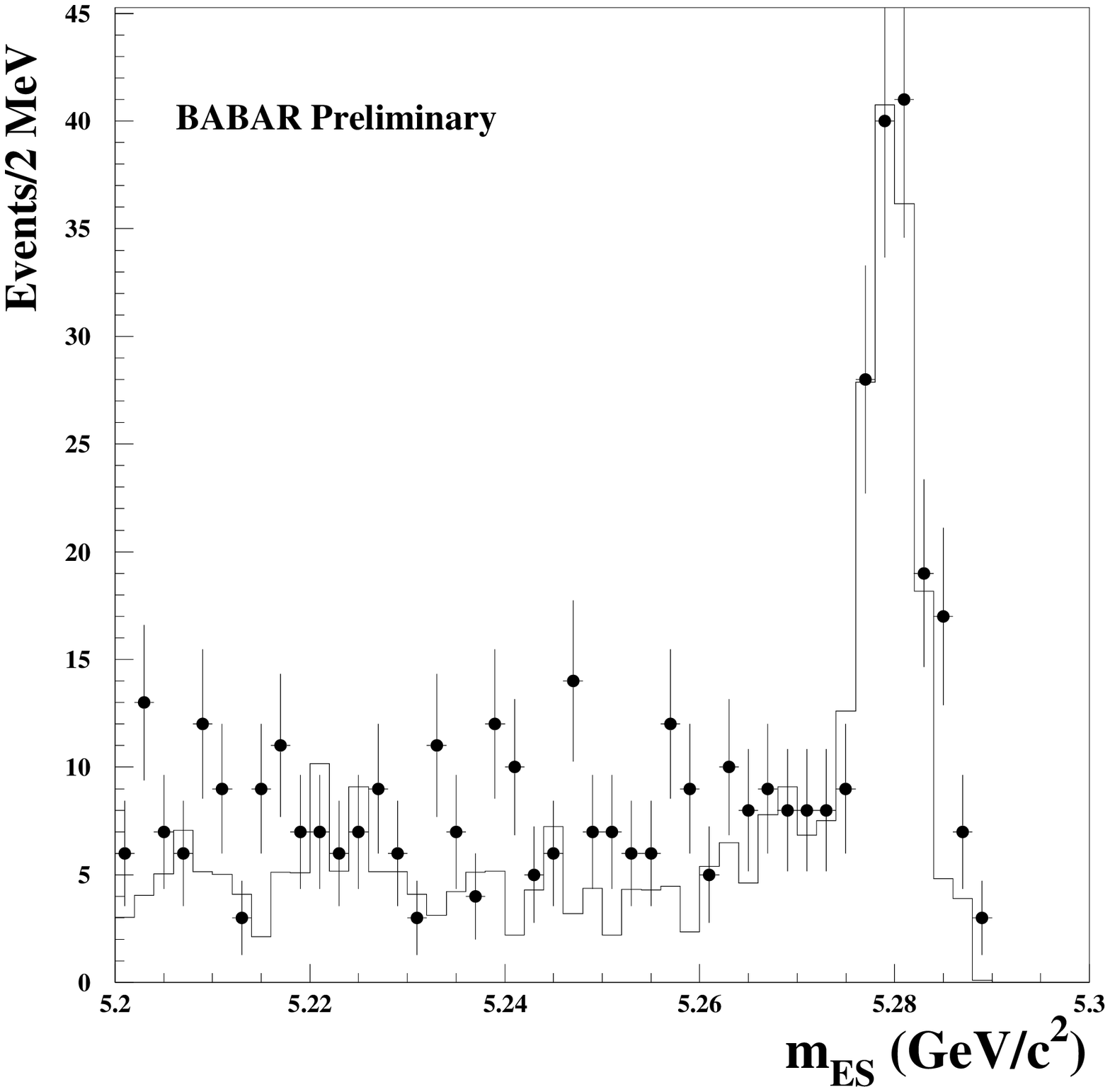}
  \includegraphics[width=0.32\textwidth]{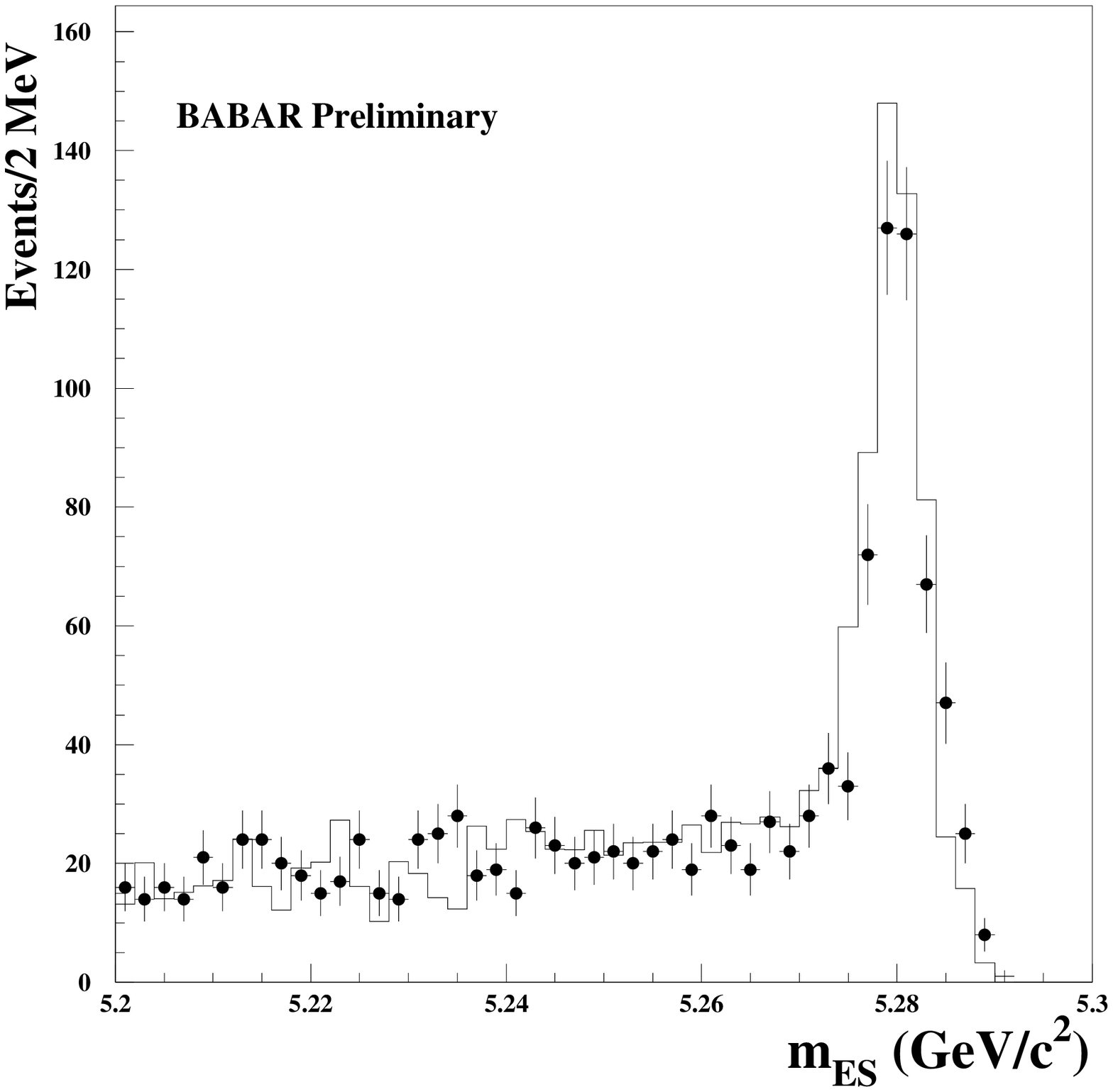}
  \includegraphics[width=0.32\textwidth]{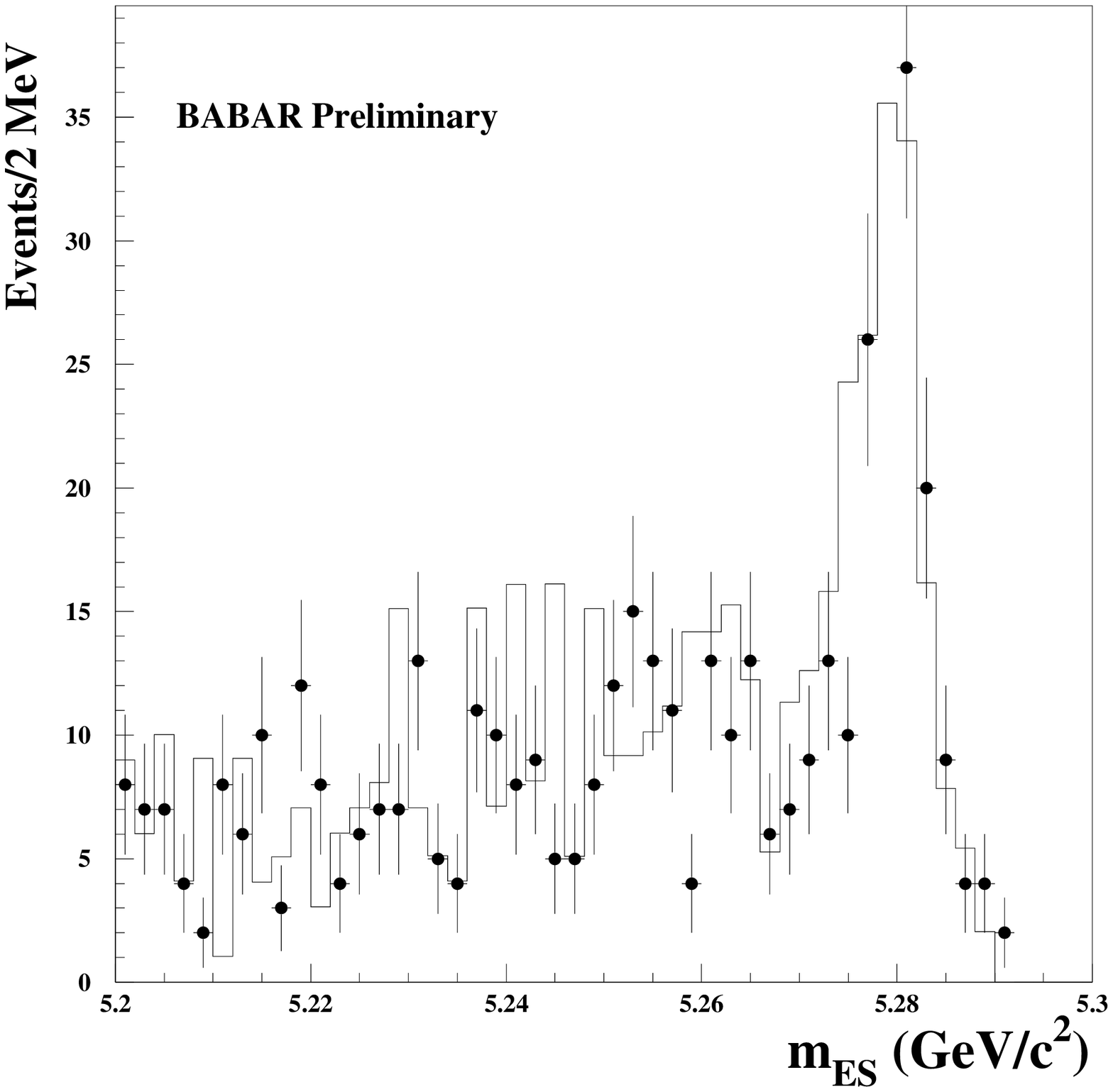}
  \includegraphics[width=0.32\textwidth]{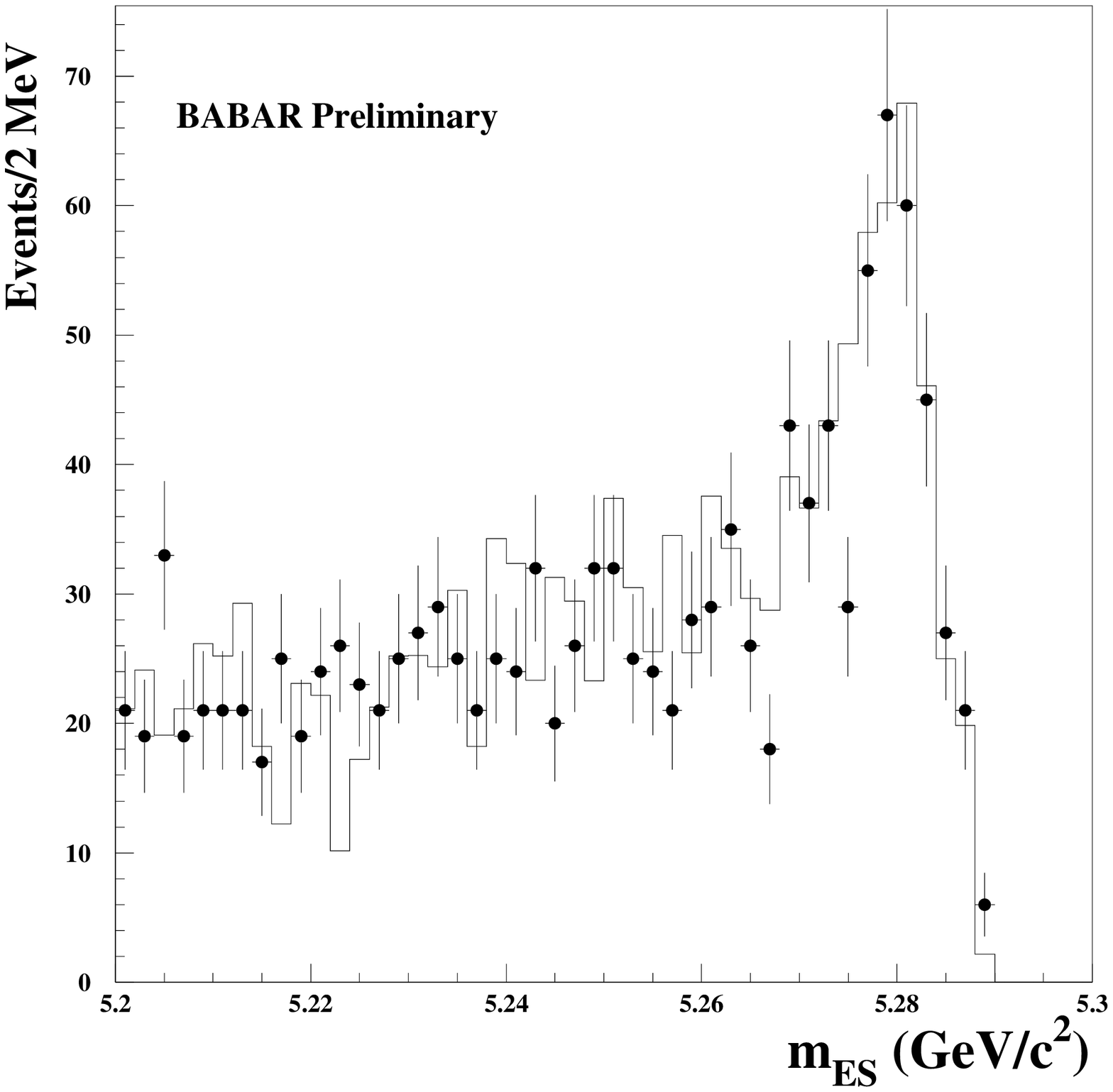}
\caption{\label{data-mc-compar} \mes distributions within the \DeltaE signal 
region. From top to bottom the rows represent the \jpsi, \psitwos and \chic1 channel. 
From left to right, the columns represent the $K^{*0}(K^+ \pi^-)$, 
$K^{*+}(\KS \pi^+)$ and $K^{*+}(K^+ \pi^0)$ channels. The points represent the data and 
the histograms represent the Monte Carlo.}
\end{center}
\end{figure}

The \B candidates, reconstructed by combining charmonium and \Kstar
candidates, are characterized by two kinematic variables: the difference
between the reconstructed energy of the \B candidate and the beam energy in the 
center-of-mass frame $\DeltaE = E_B^*-E_{beam}^*$, and the beam energy-substituted 
mass \mes, defined as $\mes \equiv \sqrt{E_{beam}^{*2}-{\bf p}_B^{*2}}$, 
where the asterisk refers to quantities in the center-of-mass and ${\bf p}_B$ is the \B momentum. 
For a correctly reconstructed \B meson, \DeltaE is expected to peak at zero and the energy-substituted mass \mes  at the \B meson mass, 5.279 \gevcc. Only one reconstructed \B meson is  allowed per event. 
For events that have multiple candidates, the candidate having 
the smallest $|\DeltaE|$  is chosen. The analysis is performed 
in a region of the \mes vs \DeltaE plane defined by $5.2 < \mes < 5.3$ \gevcc and 
$-120 < \DeltaE < 120$ \mev.
The signal region is defined as $\mes > 5.27$ \gevcc and  $|\DeltaE|$
smaller than 40 or 30 \mev for channels with or without a \piz
respectively.
Figure \ref{data-mc-compar} shows the \mes distributions for both data and Monte Carlo, within 
the \DeltaE signal region.

\section{\boldmath Angular Analysis}
\label{sec:AngularAnalysis}

The \B decay amplitudes are measured from the differential decay
distribution, expressed in the transversity basis with angles
$(\theta_{K^*},\theta_{tr},\phi_{tr})$ defined  as follows
(\cite{Aubert:2004cp,Aubert:2001pe}, Fig.~\ref{fig:heli-trans-frame})
\footnote{The conventions regarding these frame definitions are 
detailed in~\cite{Stephane,note-515}.}:
\begin{figure}[h]
\begin{center}
\includegraphics[width=0.60\linewidth]{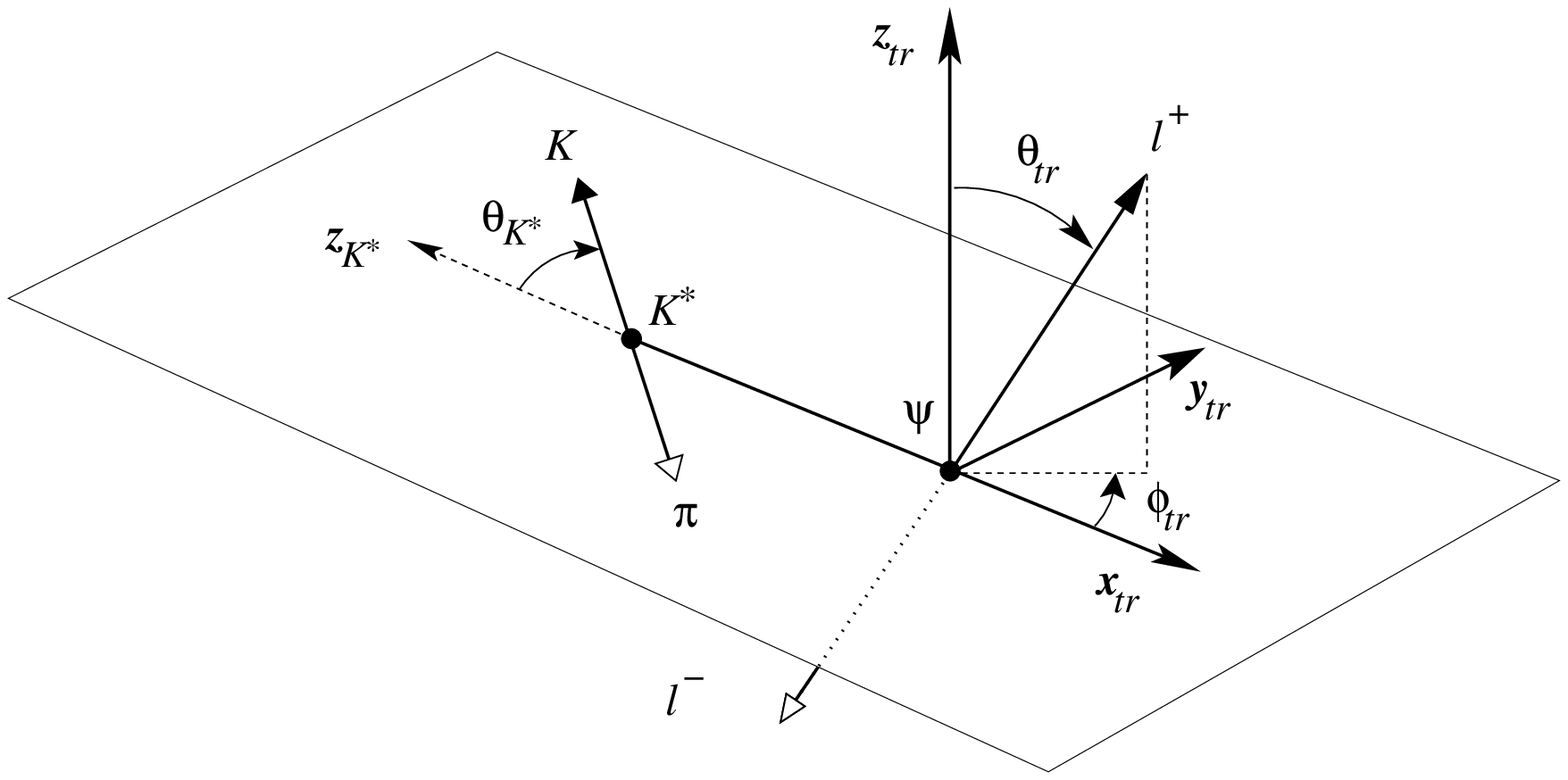}
\caption{\label{fig:heli-trans-frame}Definition of the  transversity  angles. Details are given in the text.}
\end{center}
\end{figure}
\begin{itemize}
\item $\theta_{K^*}$ is the helicity angle of the \Kstar decay. It is defined in 
the rest frame of the \Kstar meson, and is the angle between the kaon 
and the opposite direction of the $B$ meson in this frame;
\item $\theta_{tr}$ and $\phi_{tr}$ are defined in the $\Psi$ (\chicone) rest frame and are the polar and azimutal angle of 
the positive lepton (\jpsi daughter of \chicone) , with respect the axis defined by:
\begin{itemize}
\item $\boldsymbol{x}_{tr}$: opposite direction of the $B$ meson;
\item $\boldsymbol{y}_{tr}$: perpendicular to $\boldsymbol{x}_{tr}$, in the 
$(\boldsymbol{x}_{tr},\boldsymbol{p}_{K^*})$ plane, with a direction such that
$\boldsymbol{p}_{K^*}\cdot \boldsymbol{y}_{tr} > 0$;
\item $\boldsymbol{z}_{tr}$: to complete the frame, ie: 
$\boldsymbol{z}_{tr} = \boldsymbol{x}_{tr}\times\boldsymbol{y}_{tr}$.
\end{itemize}
\end{itemize}
In terms of  the angular variables
$\vomega \equiv (\theta_{K^*},\theta_{tr},\phi_{tr})$,
the  time-integrated differential decay rate for the decay of the $B$ meson  is
\begin{equation}
 \label{eqn:g_definition}
 g(\vomega;\boldsymbol{A}) \equiv 
\frac{1}{\Gamma}\frac{{\rm d}^3\Gamma}{{\rm d}\cos\theta_{K^*}{\rm d}\cos\theta_{tr}{\rm d}\phi_{tr}}=
\sum_{i=1}^{6} {\cal A}_i f_i(\vomega),
\end{equation}
where the amplitude coefficients ${\cal A}_i$ and the angular
functions $f_i(\vomega)$, $i=1\cdots 6$ are listed in Table
\ref{tab:diff:decay:rate}.
The  $\Psi$ decays to two spin-1/2 particles, while the \chicone decays to two vector particles.
The  angular
functions obtained are therefore different \cite{Stephane}.

\begin{table}[t]
\caption{\label{tab:diff:decay:rate}
Amplitude coefficients ${\cal A}_i$ and  angular functions
$f_i(\vomega)$, that contribute to the differential decay rate of a \B
meson.
An overall normalization factor $9/32\pi$ (for $\Psi$) and $9/64\pi$
(for $\chicone$) has been omitted. 
In the case of a \Bb decay, the  $\Imm$ terms change sign. }
\begin{center}
\begin{tabular}{lcccccccccc} 
\hline
\hline
$i$  &   ${\cal A}_i$               & $f_i(\vomega)$ for $\Psi$ \cite{Aubert:2004cp,Aubert:2001pe} & $f_i(\vomega)$ for \chicone \cite{Stephane}  \\ 
\hline
\noalign{\vskip1pt}
1 & $|A_0|^2                     $ & $2\cos^2\theta_{K^*}\left[1-\sin^2\theta_{tr}\cos^2\phi_{tr}\right] $ & $2 \cos^2{\theta_{K^*}} \left[1+\sin^2{\theta_{tr}}\cos^2{\phi_{tr}}\right]$ \\
2 & $|A_\parallel |^2            $ & $\sin^2\theta_{K^*}\left[1-\sin^2\theta_{tr}\sin^2\phi_{tr}\right] $ & $\sin^2{\theta_{K^*}} \left[1+\sin^2{\theta_{tr}}\sin^2{\phi_{tr}}\right]$ \\
3 & $|A_\perp|^2                 $ & $\sin^2\theta_{K^*}\sin^2\theta_{tr} $ & $\sin^2{\theta_{K^*}}\left[2\cos^2{\theta_{tr}}+ \sin^2{\theta_{tr}}\right]$ \\
4 & $\Imm (A_\parallel ^*A_\perp)$ & $\sin^2\theta_{K^*}\sin2\theta_{tr}\sin\phi_{tr} $ & $-\sin^2{\theta_{K^*}}\sin{2\theta_{tr}}\sin{\phi_{tr}}$ \\
5 & $\Ree (A_\parallel A_0^*)    $ & $-\frac{1}{\sqrt{2}}\sin2\theta_{K^*}\sin^2\theta_{tr}\sin2\phi_{tr} $ & $\frac{1}{\sqrt{2}}\sin{2\theta_{K^*}} \sin^2{\theta_{tr}} \sin{2\phi_{tr}}$ \\
6 & $\Imm (A_\perp A_0^*)        $ & $ \frac{1}{\sqrt{2}}\sin2\theta_{K^*}\sin2\theta_{tr}\cos\phi_{tr} $ & $-\frac{1}{\sqrt{2}}\sin{2\theta_{K^*}}\sin{2\theta_{tr}}\cos{\phi_{tr}}$ \\
\hline
\hline
\end{tabular}
\end{center}
\end{table}
The symbol $\boldsymbol{A}$ denotes the transversity amplitudes for the decay of the \B\ meson:
$\boldsymbol{A} \equiv (A_0,A_\parallel ,A_\perp)$.
%We set $|A_0|^2 + |A_\parallel |^2 + |A_\perp|^2 = 1$, so that $g(\vomega;\boldsymbol{A})$~(Eq.~(\ref{eqn:g_definition})) is a probability density function (PDF).
We denote by $\boldsymbol{\overline{A}}$ the amplitudes for the
$\Bbar$ meson decay. In the absence of direct \CP violation, we
can choose a phase convention in which these amplitudes are related by
$\overline{A}_0 = +A_0 $,
$\overline{A}_\parallel = +A_\parallel $,
$\overline{A}_\perp = -A_\perp $,
so that $A_\perp $ is \CP-odd and $A_0$ and ${A}_\parallel $ are \CP-even.
%Fixing this phase convention also fixes the phase of the amplitude for \Bz--\Bzb mixing.
%
The phases $\delta_j$ of the amplitudes, where $j= 0, \parallel , \perp$, are
defined by $A_j = |A_j| e^{i \delta_j}$.
Phases are defined relative to $\delta_0=0$.

\section{\boldmath Acceptance Correction}
\label{sec:AcceptanceCorrection}

We perform an unbinned likelihood fit of the three-dimensional angle PDF.
The acceptance of the detector and the efficiency of the event
reconstruction may vary as a function of the transversity angles, in
particular as the angle $\theta_{K^*}$ is strongly correlated with the
momentum of the final kaon and pion.
The PDF of the observed events, $g^{obs}$, is :
\begin{equation}
\label{eq:gObsDef}
g^{obs}(\vomega;\boldsymbol{A}) = g(\vomega;\boldsymbol{A}) 
 \frac{\varepsilon({\vomega})}{\langle \varepsilon\rangle(\boldsymbol{A})},
\end{equation}
where $g(\vomega;\boldsymbol{A})$ is given by Eq.~(\ref{eqn:g_definition}), $\varepsilon(\vomega)$ is the angle-dependent acceptance, and 
\begin{equation}
\langle \varepsilon\rangle(\boldsymbol{A}) \equiv \int{g(\vomega;\boldsymbol{A})\varepsilon(\vomega){\rm d}\vomega}
\end{equation}
is the average  acceptance. 
%over the event-weighted phase space, which depends on the amplitudes $\boldsymbol{A}$.
%
The presence of cross-feed from the companion channels which have, as
a consequence of isospin symmetry, the same $\boldsymbol{A}$
dependence as that of the signal, is taken into account.
The observed PDF for channel $\jchan$ $(\jchan=\Kpm\pimp, \KS\pipm, \Kpm\piz)$ is then
\begin{eqnarray}
\label{eq:gobs}
g^{\jchan,obs}(\vomega;\boldsymbol{A}) &=&
g(\vomega;\boldsymbol{A})
\frac{\varepsilon^{\jchan}(\vomega)}
{\sum_{k=1}^{6}{\cal A}_k(\boldsymbol{A}) \Phi^{\jchan}_k}
\end{eqnarray}
$\varepsilon^{\jchan}(\vomega)$ is the efficiency for
reconstructed channel $\jchan$ considering $\B\to (c\bar{c}) \Kstar$ channels as a 
whole (for the three charmonium states separately), that is counting cross-feed events as signal.
The   $\Phi^{\jchan}_k$ are the $f_k(\vomega)$ moments of the
``whole'' efficiency $\varepsilon^{\jchan}$.
The expressions for 
$\varepsilon^{\jchan}(\vomega)$ 
and  $\Phi^{\jchan}_k$
are available and discussed in section  IV.A of Ref. \cite{Aubert:2004cp}.

The acceptance $\varepsilon^{\jchan}(\vomega)$ can be expressed as in
Eq. (\ref{eq:gobs}), and only the coefficients $\Phi^{\jchan}_k$ are
needed, under the approximations that the angular resolution can be
neglected, including for cross-feed events, and that the double
misidentification of the daughters of the $\Kstarz \to \Kpm \pimp$
candidate ($K$--$\pi$ swap) can be neglected.
The biases induced by these approximations have been estimated with   Monte Carlo
(MC) based studies, and found to be negligible (see table IV of \cite{Aubert:2004cp}).

The coefficients $\Phi^{\jchan}_k$ are computed with exclusive
signal MC samples obtained using a full simulation of the experiment
\cite{geant,Evtgen}.
PID efficiencies measured with data control samples
are used to adjust the MC simulation to the actual behavior of the detector.
Separate coefficients are used for 
different charges of the final state mesons, in particular to take into account
the charge dependence of the interaction of charged kaons
with matter, and a possible charge asymmetry of the detector.
Writing the expression for the  log-likelihood 
$L^{\jchan}(\boldsymbol{A})$
for the PDF 
$g^{\jchan,obs}(\vomega_i;\boldsymbol{A})$
for a pure signal sample of $N_{S}$ events, 
the relevant contribution is 
\begin{equation} 
\label{eqn:likelihoodDef}
L^{\jchan}(\boldsymbol{A})=
\sum_{i=1}^{N_{S}}\ln\left( g(\vomega_i;\boldsymbol{A}) \right) - N_{S} \ln\left(\sum_{k}{\cal A}_k(\boldsymbol{A}) \Phi^{\jchan}_k\right), 
\end{equation}
since the remaining term 
$ \sum_{i=1}^{N_{S}}\ln\left(\varepsilon^{\jchan}(\vomega_i)\right)$
does not depend on the amplitudes.

\section{\boldmath Background Correction}
\label{sec:BackgroundCorrection}

We use a background correction method described in section  IV.B of Ref. \cite{Aubert:2004cp}, 
in which background events are added with a negative weight to the log-likelihood that is 
maximized

\begin{equation}
\label{eqn:pseudo-log}
 L^{\jchan\prime}(\boldsymbol{A}) \equiv \sum_{i = 1}^{n_{B}+N_{S}}\ln(g^{\jchan,obs}(\vomega_i;\boldsymbol{A}))
- \frac{\tilde{n}_B}{N_{B}}\sum_{k = 1}^{N_{B}}\ln(g^{\jchan,obs}(\vomega_k;\boldsymbol{A})).
\end{equation}

The fit is performed within the \mes signal region ($\mes > 5.27\gevcc$) which contains $N_B$ signal 
events. $\tilde{n}_B$ is an estimate of the unknown number $n_B$ of background events that are present 
in the signal region in the data sample. In Ref. \cite{Aubert:2004cp} 
background events were estimated from the data by fitting the \mes distribution in the sideband 
region and extrapolating in the \mes signal region. This method assumed that the background 
has only a combinatorial contribution and no peaking contribution. This is a valid argument for 
the \jpsi channels, but not the \psitwos and \chicone channels where peaking backgrounds are known 
to have a non-negligigle contribution. Therefore, in this analysis, the background (combinatorial 
and peaking) has been taken from generic MC. As $L^{\jchan\prime}$ is not a log-likelihood, the uncertainties 
yielded by the minimization program \minuit~\cite{minuit} are biased estimates of the actual uncertainties. 
An unbiased estimation of the uncertainties is described and  validated in Appendix~A of 
Ref. \cite{Aubert:2004cp}. With this pseudo-log-likelihood technique, we avoid parametrizing the 
acceptance as well as the background angular distributions.

\section{Systematics}
\label{sec:Systematics}

The measurement is affected by several systematic uncertainties.
The branching ratios that are used in the cross-feed part of the acceptance cross section are varied by 
$\pm 1 \sigma$, and the largest variation  is retained.
The uncertainty induced by
the finite size of the MC sample used to compute  the coefficients $\Phi_k^b$  is estimated by the statistical uncertainty of the angular fit on that MC sample (shift additivity \cite{Aubert:2001pe}).
The uncertainty due to our limited understanding of the PID is
estimated by using two different methods to correct for the MC-vs-data
differences.
The background uncertainty is obtained by comparing MC and data
shapes of the \mes distributions for the combinatorial component
and by using the corresponding branching errors for the peaking
component.
The uncertainty due to the presence of a $K\pi$ S wave under the
\Kstarone peak is estimated by a fit including it.
The differential decay rate is described by eqs. (6--9) of 
 Reference \cite{Aubert:2004cp}.

\section{Results}

\begin{figure}[ht]
\begin{center}
\includegraphics[width=0.48\linewidth]{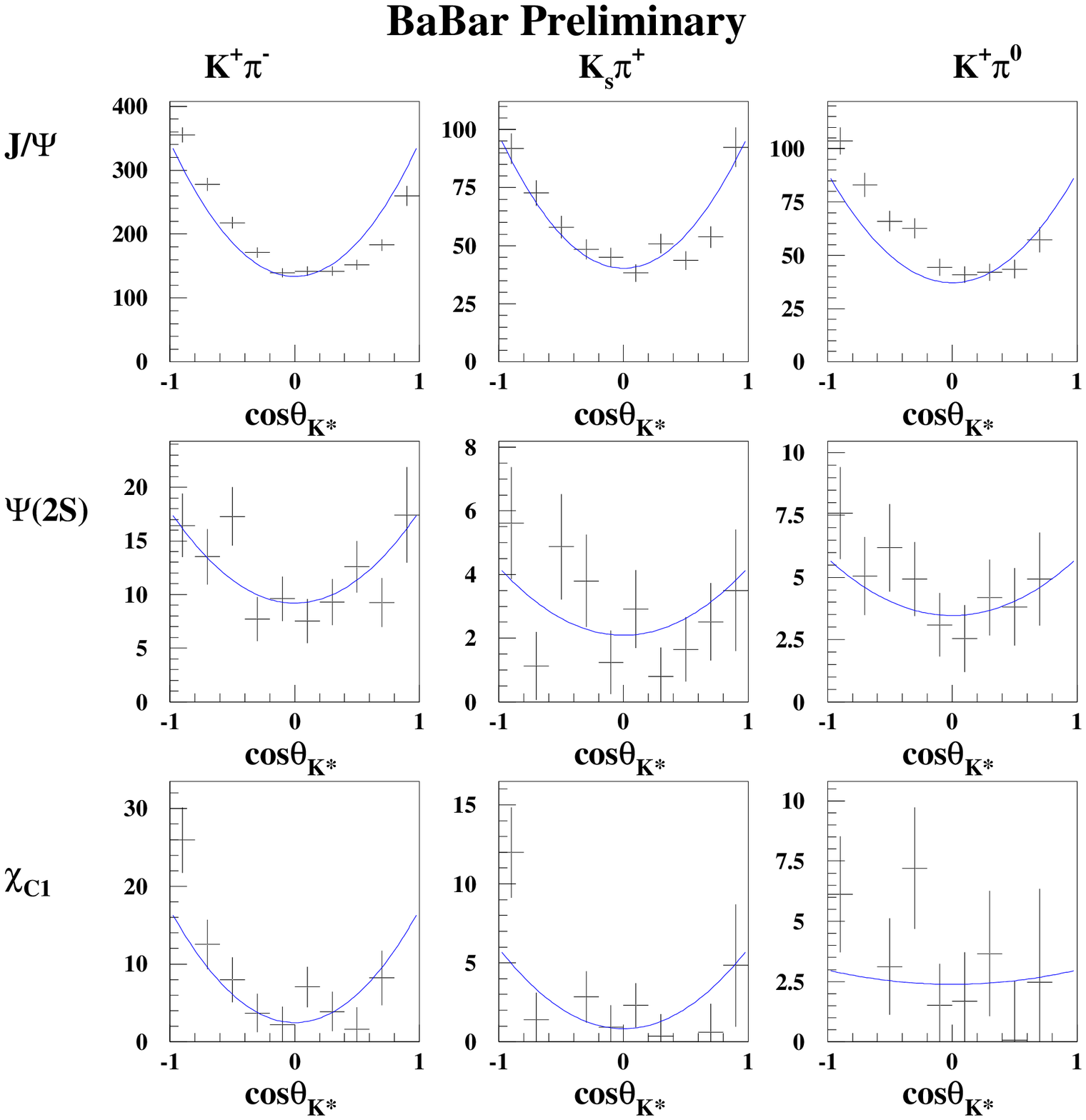}
\includegraphics[width=0.48\linewidth]{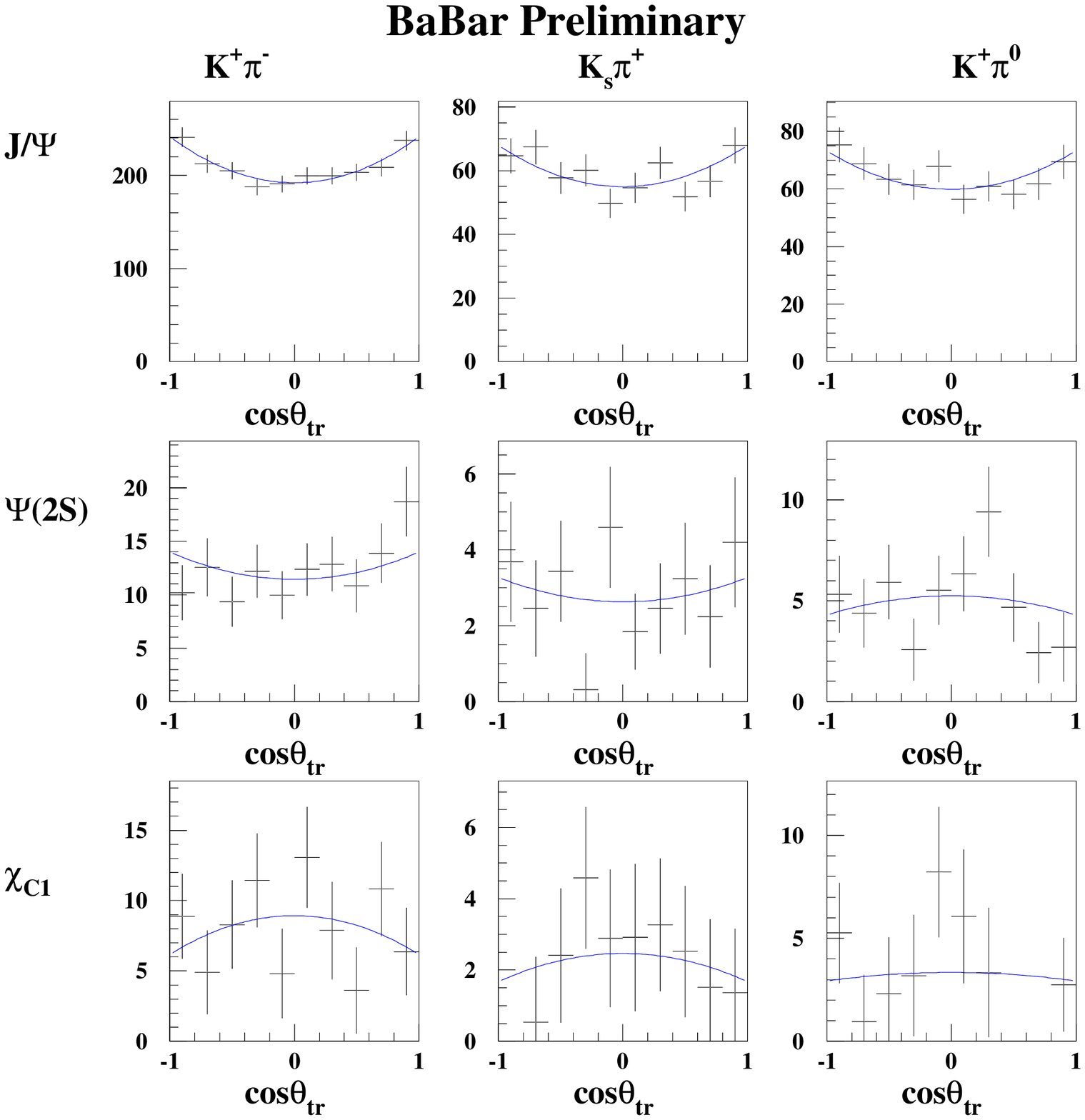}
\includegraphics[width=0.48\linewidth]{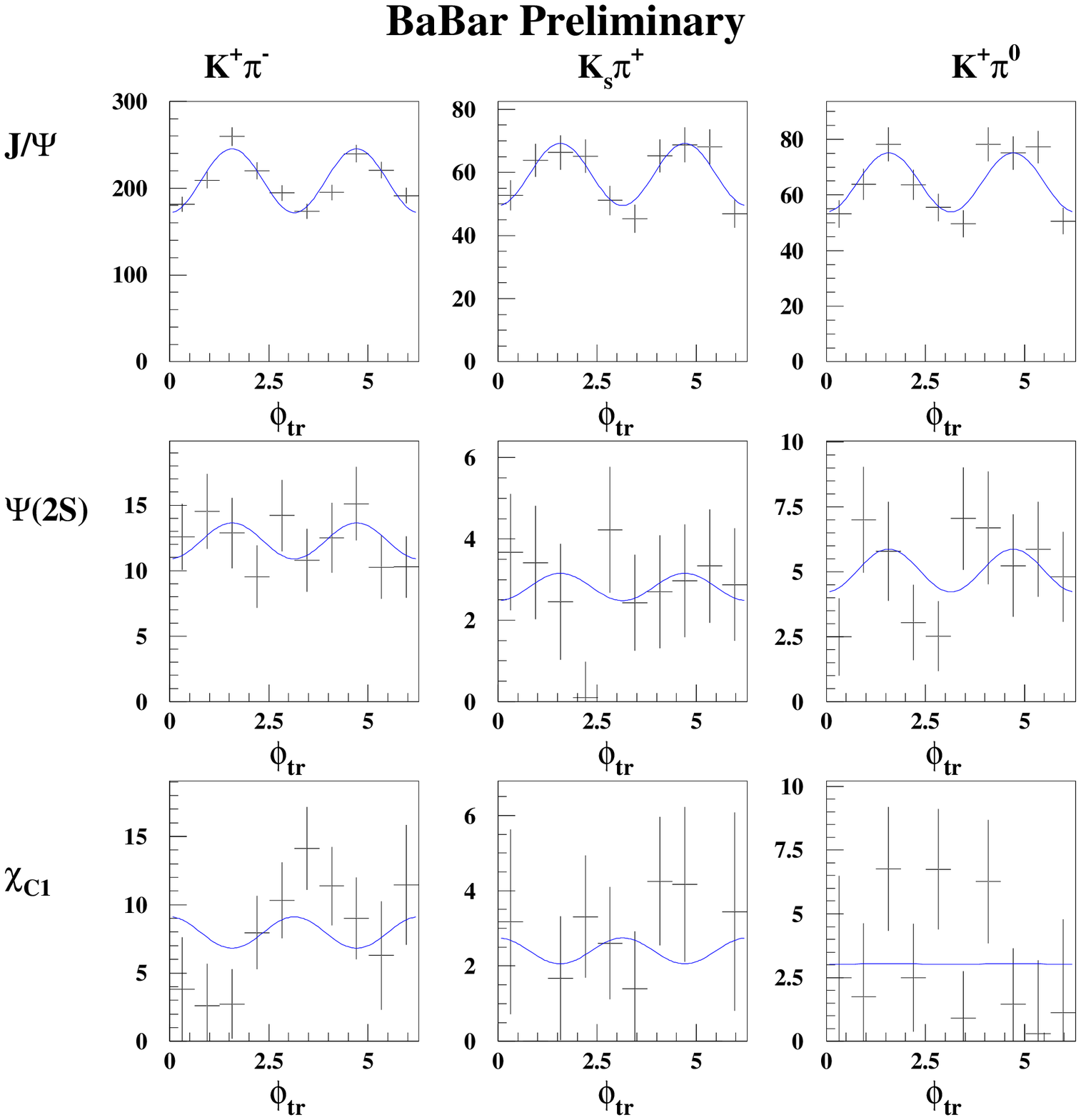}
\caption{\label{fig:resultat}
Angular distributions with PDF from fit overlaid.
The asymmetry of the  $\cos\theta_{K^*}$ distributions induced by the S-wave inteference 
is clearly visible.}
\end{center}
\end{figure}

The results are summarized in Table~\ref{tab-res1}.
The values of $|A_0|^2$, $|A_\parallel|^2$, $|A_\perp|^2$ turn out to
be negatively correlated, as is expected for quantities the sum of
which is unity. In particular, $|A_\parallel|^2$, which would be the
least precisely measured parameter in separate one-dimensional fits, is strongly
anti-correlated with $|A_0|^2$, which would be the best measured.
The one-dimensional (1D) distributions,
acceptance-corrected with an 1D Ansatz
\footnote{
In contrast with the dedicated method used in the fit, 
for the plots, we simply computed the 1D efficiency maps from the
distributions of the accepted events divided by the 1D PDF.}
, and background-subtracted, are 
 overlaid with the fit results and shown on figure 
\ref{fig:resultat}.
As in lower statistics studies, the $\cos\theta_{K^*}$ forward backward asymmetry due to the interference with the S wave is clearly visible.

\begin{figure}[ht]
\begin{center}
\includegraphics[width=0.49\linewidth]{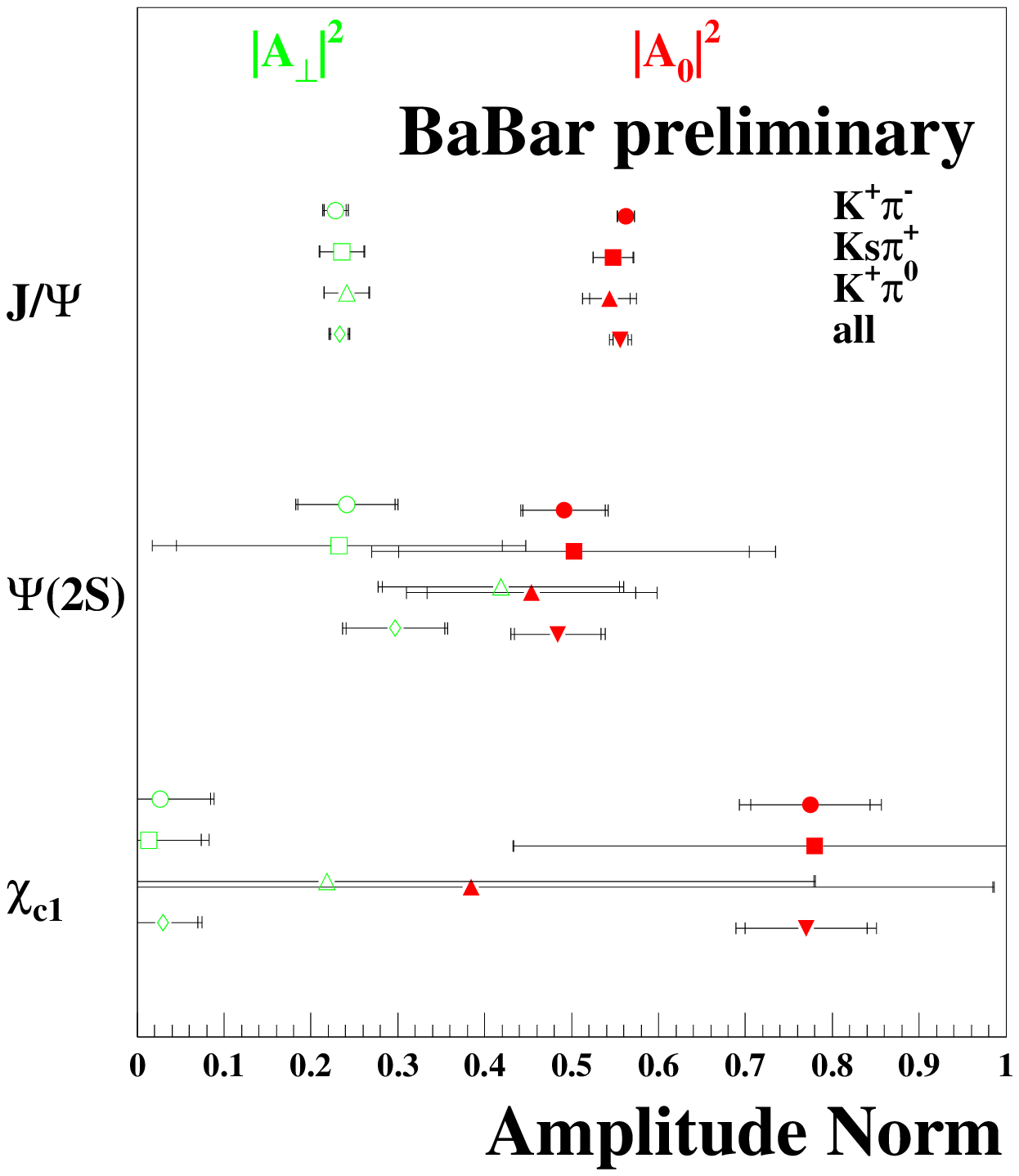}
\includegraphics[width=0.49\linewidth]{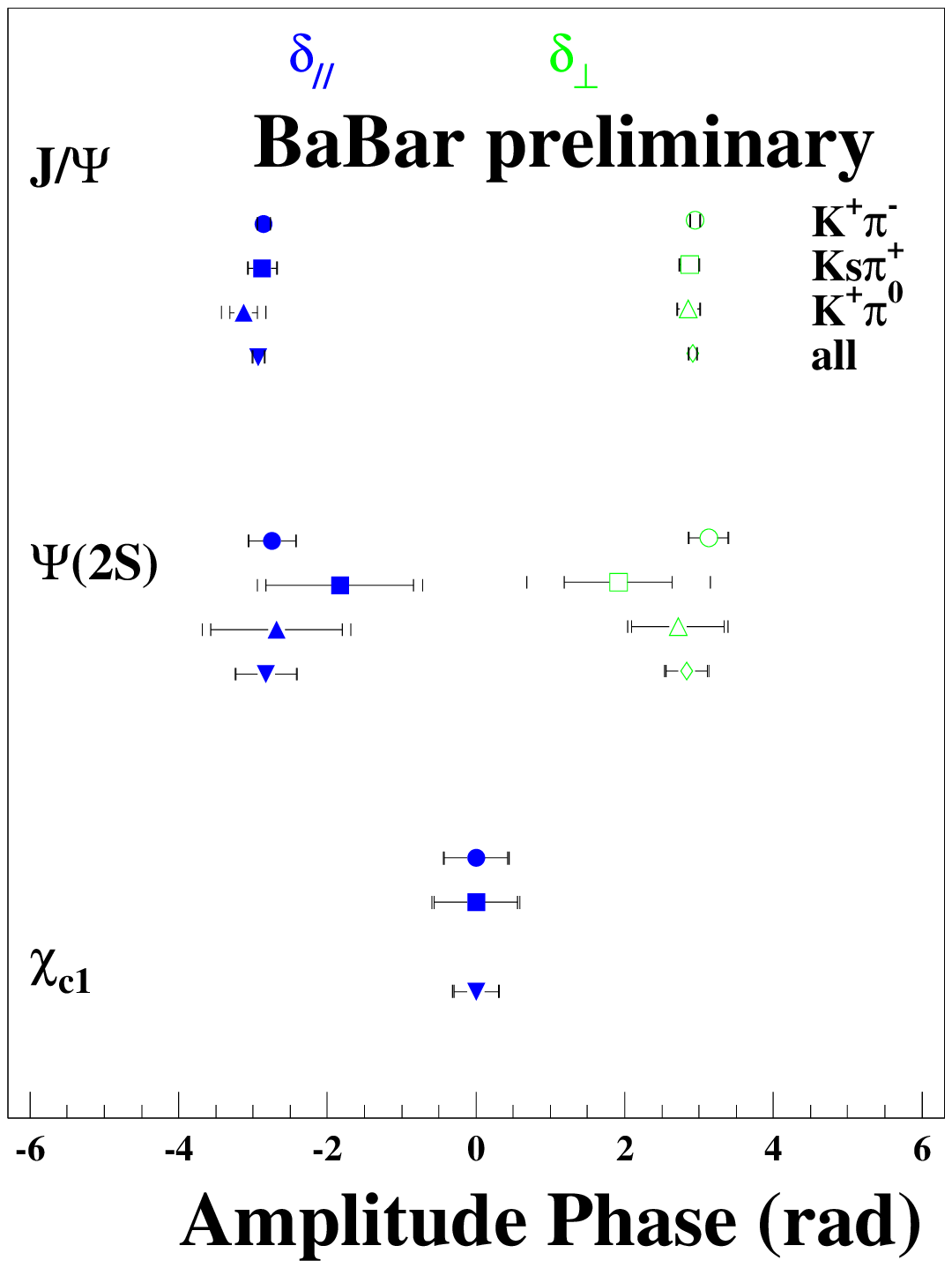}
\caption{\label{fig:results:summary} 
Data : Results of the fits. 
Left  : norms ($|A_0|^2$ in red,  $|A_\parallel|^2$ in green), 
Right  : phases ($\delta_\parallel$ in blue,  $\delta_\perp$  in green).
Circles :\Kpm\pimp; Squares : \KS\pip; Up triangles : \Kpm\piz,  Down triangles  : all \Kstar's combined.
In the case of decays to \chicone,  $A_\perp$ is compatible with zero, and therefore its phase is not defined. 
(the $\Kp\piz$ contribution is removed in computing $\delta_\parallel$ for \chicone, as   $|A_\parallel|^2$ is compatible with zero).
For each charmonium and each measurement, the results for the individual channels 
are shown together with the average.}
\end{center}
\end{figure}

\begin{table}[h] \footnotesize
\caption{\label{tab-res1}Summary of the amplitudes measured.
In the case of decays to \chicone,  $A_\perp$ is compatible with zero, and therefore its phase is not defined.}
\begin{center}\scriptsize
\begin{tabular}{ccccccc} \hline \hline
\noalign{\vskip1pt}
Channel & $|A_0|^2$ & $|A_\parallel|^2$ &  $|A_\perp|^2$  & $\delta_\parallel$ &  $\delta_\perp$  \\ \hline
$J/{\psi} K^*$ & $0.556\pm0.009\pm0.010$ & $0.211\pm0.010\pm0.006$ & $0.233\pm0.010\pm0.005$  & $-2.93 \pm 0.08 \pm 0.04$  &  $2.91 \pm 0.05 \pm 0.03 $ \\  \hline
$\psi(2S) K^*$ & $0.48\pm0.05\pm0.02$ & $0.22\pm0.06\pm0.02$ & $0.30\pm0.06\pm0.02$ & $-2.8 \pm 0.4 \pm 0.1$  &  $2.8 \pm 0.3 \pm 0.1 $  \\  \hline
$\chi_{c1} K^*$ & $0.77\pm0.07\pm0.04$ & $0.20\pm0.07\pm0.04$ & $0.03\pm0.04\pm0.02$ & $0.0 \pm 0.3 \pm 0.1$  &  --  \\  \hline \hline
\end{tabular}
\end{center}
\end{table}
A graphical representation is given in Fig. \ref{fig:results:summary}.

\begin{table} 
\caption{\label{tab-res:dir} 
Difference between the interference terms measured in \B and \Bb decays to \jpsi.
}
\begin{center}
\begin{tabular}{c|ccc}
\hline\hline
 & $(K^+ \pi^-)$ & $(K^+ \pi^0)$ & $(\KS \pi^+)$ \\ \hline
$\delta {\cal A}_4$ & $ 0.002 \pom  0.025 \pom 0.005 $ & $ -0.017 \pom  0.047 \pom 0.023 $ & $  -0.008 \pom  0.049 \pom 0.011$ \\
$\delta {\cal A}_6$ & $  -0.011 \pom  0.043 \pom 0.016 $ & $  -0.051 \pom  0.098 \pom 0.064 $ & $ 0.075 \pom  0.089 \pom 0.009$\\
\hline\hline
\end{tabular}
\end{center}
\end{table} 

\clearpage

In summary,
\begin{itemize}
\item
Our measurement of the amplitudes of \B decays to \jpsi are compatible
with, and of better precision than, previous measurements.
\item
From a comparison of \Bz and \Bp decays,
 isospin is seen to be conserved in the decay.
\item
We confirm our previous observation that the strong phase differences
are significantly different from zero, in contrast with what is predicted by factorization.
For $\B\to\jpsi\Kstar$, it amounts to 
$\delta_\parallel - \delta_\perp = 0.45 \pm 0.05 \pm 0.02$, an $8 \sigma$ effect.
\item
The presence of direct \CP-violating triple-products in the amplitude
would produce a \B to \Bb difference in the interference terms ${\cal A}_4$ and ${\cal A}_6$: 
$\delta {\cal A}_4$ and $\delta {\cal A}_6$. This is not observed 
as all the measurement are compatible with zero (see Table~\ref{tab-res:dir}), 
with an improved precision with respect to the BELLE measurement \cite{belletp}.
\item
We have performed the first three-dimensional analysis of the decays to \psitwos
and \chicone. 
The longitudinal polarization of the decay to \psitwos is
smaller than that of the \jpsi, 
while the \CP-odd intensity fraction of the two decays are similar. This is
 compatible with the
prediction of models of meson decays in the framework of
factorization.

The longitudinal polarization of the decay to \chicone is found to be
larger than that to \jpsi, in contrast with the predictions of
Ref. \cite{Chen:2005ht} which include non-factorizable contributions.
The \CP-odd intensity fraction of this decay is 
compatible with zero.
The phases of the parallel and of the longitudinal amplitudes are
observed to be compatible with each other, in contrast with decays to
$\Psi$.
\end{itemize}

We are grateful for the 
extraordinary contributions of our \pep2\ colleagues in
achieving the excellent luminosity and machine conditions
that have made this work possible.
The success of this project also relies critically on the 
expertise and dedication of the computing organizations that 
support \babar.
The collaborating institutions wish to thank 
SLAC for its support and the kind hospitality extended to them. 
This work is supported by the
US Department of Energy
and National Science Foundation, the
Natural Sciences and Engineering Research Council (Canada),
Institute of High Energy Physics (China), the
Commissariat \`a l'Energie Atomique and
Institut National de Physique Nucl\'eaire et de Physique des Particules
(France), the
Bundesministerium f\"ur Bildung und Forschung and
Deutsche Forschungsgemeinschaft
(Germany), the
Istituto Nazionale di Fisica Nucleare (Italy),
the Foundation for Fundamental Research on Matter (The Netherlands),
the Research Council of Norway, the
Ministry of Science and Technology of the Russian Federation, 
Ministerio de Educaci\'on y Ciencia (Spain), and the
Particle Physics and Astronomy Research Council (United Kingdom). 
Individuals have received support from 
the Marie-Curie IEF program (European Union) and
the A. P. Sloan Foundation.

\clearpage

%%%%%%%%%%%%%%%%%%%%%%%%%%%%%%%%%%%%%%%%%%%%%%%%%%%%%%%%%%%%%%%%%%%%%%%%%%

\end{document}